\documentclass{article}

\PassOptionsToPackage{sort&compress,numbers}{natbib}

\usepackage[preprint]{neurips_2021_ml4ps}

\usepackage[utf8]{inputenc} 
\usepackage[T1]{fontenc}    
\usepackage{hyperref}       
\usepackage{url}            
\usepackage{booktabs}       
\usepackage{amsfonts}       
\usepackage{nicefrac}       
\usepackage{microtype}      
\usepackage{xcolor}         
\usepackage{amsmath}
\usepackage{graphicx}

\newcommand{\cL}{\mathcal{L}}
\newcommand{\gpred}{g_{\text{pr}}}

\newcommand{\gFS}{\tilde{g}_{\text{FS}}}
\newcommand{\gCY}{g}
\newcommand{\gFSX}{g_{\text{FS}}}

\newcommand{\vCY}{\text{Vol}_{\text{CY}}}

\renewcommand{\Re}{{\rm Re}\;} 
\renewcommand{\Im}{{\rm Im }\;} 

\title{
	Learning Size and Shape of Calabi-Yau Spaces}

\author{%
	Magdalena Larfors \\
	Department of Physics and Astronomy\\
	Uppsala University\\
	SE-751 20 Uppsala \\
	Sweden \\
	\texttt{magdalena.larfors@physics.uu.se} \\
	\AND
	Andre Lukas \\
	Rudolf Peierls Centre for Theoretical Physics\\
	Oxford University\\
	Parks Road, Oxford OX1 3PU\\
	United Kingdom \\
	\texttt{andre.lukas@physics.ox.ac.uk} \\
	\AND
	Fabian Ruehle \\
	Department of Physics\\
	Northeastern University\\
	360 Huntington Avenue, Boston MA 02115\\
	United States\\
	\texttt{f.ruehle@northeastern.edu} \\
	\AND
	Robin Schneider \\
	Department of Physics and Astronomy\\
	Uppsala University\\
	SE-751 20 Uppsala \\
	Sweden \\
	\texttt{robin.schneider@physics.uu.se} \\
}

\begin{document}
	
	\maketitle
	
	\begin{abstract} 
		We present a new machine learning library for computing metrics of string compactification spaces. We benchmark the performance on Monte-Carlo sampled integrals against previous numerical approximations and find that our neural networks are more sample- and computation-efficient. We are the first to provide the possibility to compute these metrics for arbitrary, user-specified shape \textit{and} size parameters of the compact space and observe a linear relation between optimization of the partial differential equation we are training against and vanishing Ricci curvature.   
	\end{abstract}
	
	\section{Introduction}
	\label{sec:intro}

	String theory is the leading candidate for a unification of general relativity and the standard model of particle physics. Its mathematical consistency, however, requires the introduction of six additional small and compact dimensions, usually described by Calabi-Yau (CY) manifolds. The shape, size, and topology of these internal spaces determine many interesting physical quantities such as the gauge group and particle content of the theory. CY manifolds are not only interesting objects in physics, but exhibit fascinating mathematical properties, such as mirror symmetry.

	CY spaces have been extensively studied, starting with~\citep{Candelas:1987kf,Green:1987cr}. However, above four compact dimensions, there exists no known closed form expression for their metric tensor. The metric is of utmost importance to connect string theory with observable physics. It is a necessary ingredient in the computation of physical Yukawa couplings from string theory and can be used to test conjectures from the swampland program~\citep{Vafa:2005ui,Ashmore:2021qdf} or regarding mirror symmetry~\citep{Strominger:1996it}. These applications furthermore require control over \textit{how} the metric depends on the size and shape parameters, so-called {\it moduli}, of the CY geometry.

	The standard numerical approximation for CY metrics is the Donaldson algorithm~\citep{2005math.....12625D}, a fixed point iteration methods which expands the metric in a monomial basis of degree $k$ and converges to the CY metric as $k\rightarrow \infty$. This algorithm has been utilized in examples~\citep{Douglas:2006rr,Braun:2007sn,Braun:2008jp,Douglas:2006hz,Anderson:2010ke,Anderson:2011ed}. However, there are three main drawbacks: first, its time and space complexity scales poorly (factorially) with $k$. Second, changing the moduli of the metric requires running the whole fixed point iteration from scratch. Lastly, the algorithm is in fact providing a metric with {\it different} properties, which happens to coincide with the CY metric in the limit $k\rightarrow \infty$. However, better direct optimizations of the CY metric are possible for any finite, fixed $k$. Energy functionals~\citep{Headrick:2009jz,Cui:2019uhy} and holomorphic neural networks~\citep{Douglas:2020hpv} have proven more efficient. Alternatively, one may directly learn the CY metric tensor~\citep{Ashmore:2019wzb,Anderson:2020hux,Jejjala:2020wcc}. By modeling the metric as neural network, this has also allowed studies of the shape moduli dependence~\citep{Anderson:2020hux}.

	In this paper we advance this promising line of research by presenting an open source package, \texttt{cymetric}\footnote{\url{https://github.com/pythoncymetric/cymetric}}, to learn CY metrics with TensorFlow~\citep{tensorflow2015-whitepaper} on a wide class of CY manifolds.  Given the lack of analytic metric solutions, direct supervised training of the neural network is impossible. We therefore proceed in an almost self-supervised manner and introduce five loss functions that govern the learning process. Our neural network reproduces the full moduli dependence of the CY metric, thus filling a crucial knowledge gap which has so far impeded phenomenological studies in string theory. The implementation improves sample and computation efficiency compared to previous approximations. Moreover, experiments run with our package establish a direct linear relation between optimizing a simple surrogate loss and the computationally much more expensive vanishing Ricci curvature.

	The results of our experiments have promising implications for utilizing neural networks in solving differential equations~\citep{jin2020unsupervised,raissi2019physics}. In particular, we demonstrate that by using physical domain knowledge, it is possible to find solutions to non-linear second-order differential (Monge-Amp\`ere) equations by computing only a single derivative in the training process. 
	There are (besides computation time) no limitations on generating points samples as input data, and each sample point comes with machine precision. This opens up the possibility to study various scaling laws of neural networks~\citep{DBLP:journals/corr/abs-2001-08361}.
	The training data consists of Monte-Carlo sampled points on the CY. Using a powerful theorem of Shiffman and Zelditch~\citep{1999CMaPh.200..661S}, we do not need to perform MCMC, but can construct the measure directly by embedding the manifold into a higher-dimensional ambient space with known measure.

	\section{Methodology}
	\label{sec:method}

	\paragraph{Calabi-Yau manifolds}

	A CY manifold, $X$, is a compact, complex, Ricci-flat K\"ahler manifold. To describe it, we recall a few items from differential, complex and K\"ahler geometry. A CY geometry is specified by two nowhere-vanishing differential forms: a real K\"ahler 2-form $J$, associated to a K\"ahler metric $\gCY$, and a holomorphic $n$-form $\Omega$, which specifies how $n$ holomorphic coordinates are selected among the $2n$ real ones. Given $X$, the K\"ahler form is not unique; however, within the same cohomology class of any closed $J'$, there is unique representative $J$ whose corresponding metric $\gCY$ has vanishing Ricci curvature by the celebrated Calabi-Yau theorem~\citep{Calabi+2015+78+89,Yau:1978cfy}. On any K\"ahler manifold, the metric is encoded in a K\"ahler potential $\mathcal{K}$, and the Ricci curvature simplifies to 
	\begin{align}
		\label{eq:ricciscalar}
		g=\partial\bar{\partial} \mathcal{K} \; ,\qquad \; R =  \partial \bar{\partial} \log \det g \; .
	\end{align}
	The unique volume form of a CY three-fold can be expressed in terms of either $J$ or  $\Omega$. Combining this with the cohomological equivalence of $J$ and $J'$, 
	we get the {\it Monge-Amp\`ere equation},
	\begin{align}
		\label{eq:MA}
		J \wedge J \wedge J = \kappa \; \Omega \wedge \bar{\Omega} = \kappa \; \text{d} \vCY \qquad \text{ with } \qquad  J = J' + \partial \bar{\partial} \phi
	\end{align}
	for a real scalar function $\phi$, where $\kappa$ is some complex constant. This second order complex non-linear partial differential equation is satisfied iff $\gCY$ is the unique Ricci-flat metric on $X$~\cite{Calabi+2015+78+89,Yau:1978cfy}.
	
	CY manifolds are commonly constructed as (intersections of) hypersurfaces, i.e. the vanishing loci of polynomial equations, in a higher-dimensional ambient space, $\mathcal{A}$. In this paper we consider complete intersection CY manifolds in $\mathcal{A} = \Pi_{i=1}^r \mathbb{P}^{n_i}$, where $\mathbb{P}^{n_i}$ are complex projective spaces~\citep{Candelas:1987kf,Green:1987cr}.  
	Our TensorFlow models also work for the numerous CY hypersurfaces in toric varieties specified in the Kreuzer-Skarke list~\citep{Kreuzer:2000xy} and for CY manifolds in complex dimension other than three. The performance of our model is comparable for projective and toric ambient spaces, but is reduced for CY manifolds that are either defined by multiple hypersurfaces, or have many shape moduli.

	\paragraph{Point Sampling} 
	The training data consists of points on the CY $X$, which must be sampled uniformly with respect to a known measure. Rejection sampling is not feasible due to the high dimensionality of our problem. MCMC methods can deal with high dimensionality~\citep{hoffman2014no}, but they require rejecting many points, and so are not optimal. We therefore implemented a more efficient point sampling method in NumPy~\citep{harris2020array}, which employs theorems from complex geometry~\citep{1999CMaPh.200..661S,Braun:2007sn}, and applies to CY spaces constructed from intersecting hypersurfaces. This results in points sampled according to a measure $\text{d}A$, which is computed as an anti-symmetric product of certain ambient space measures pulled back to the CY. A function $f$ may then be Monte-Carlo integrated over the CY using
	\begin{align}
		\label{eq:MCint}
		\int_X \text{d} \vCY f  = \int_X \frac{\text{d} \vCY}{\text{d}A} \text{d}A\; f  = \frac{1}{N} \sum_i w_i f |_{p_i} \quad \text{ with }\quad   w_i = \frac{\text{d} \vCY}{\text{d}A} |_{p_i}\;,
	\end{align}
	where the measure d$A$ and weights $w_i$ are given by the outlined procedure. Finally, the performance of numeric metric approximations is evaluated by integrating equations~\eqref{eq:ricciscalar} and \eqref{eq:MA} over $X$ using~\eqref{eq:MCint}.
	
	\begin{figure}[t]
		\centering
		\includegraphics[width=1.\textwidth]{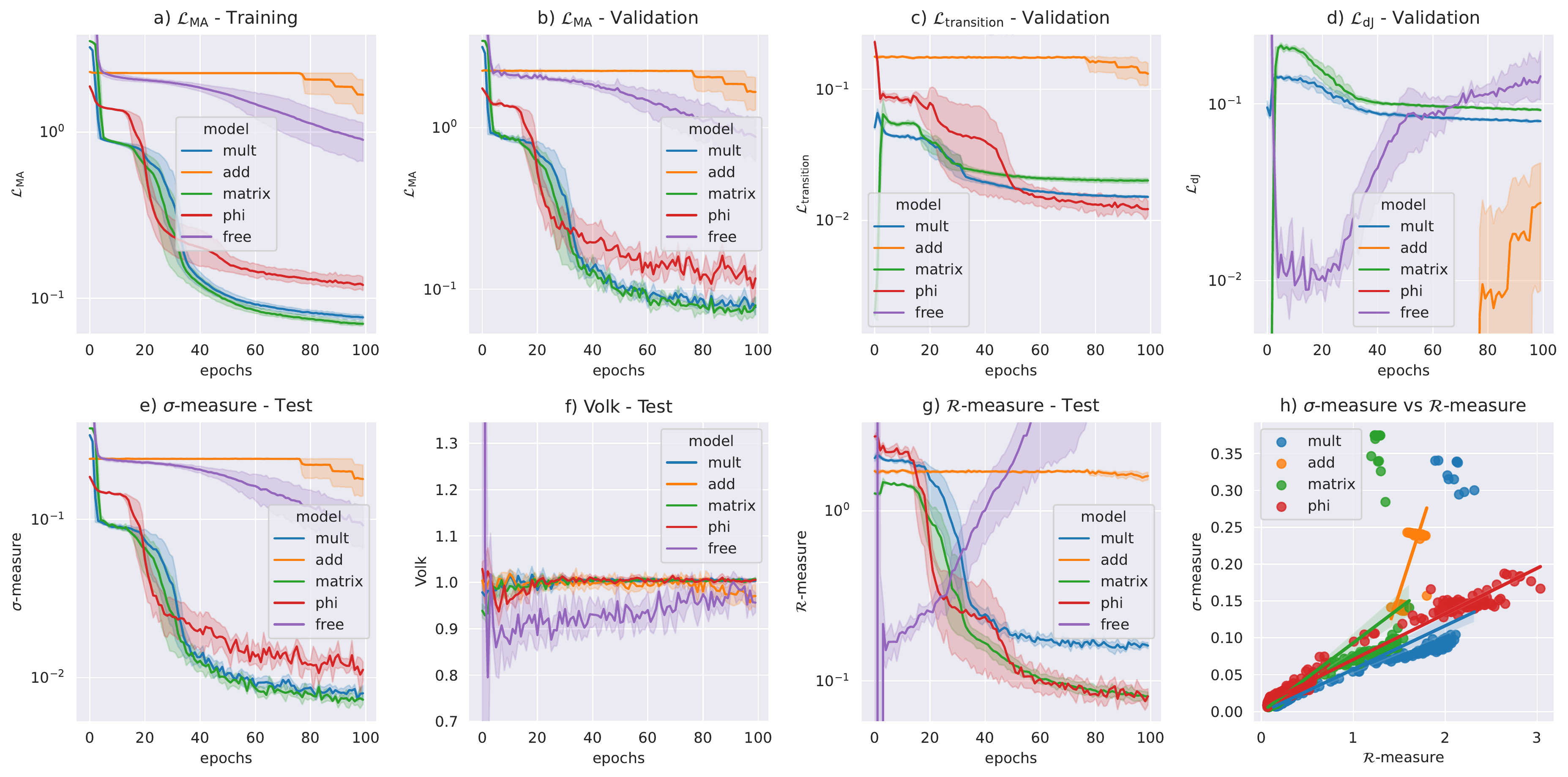}
		\caption{Fermat Quintic experiments: a) Monge-Amp\`ere loss on training data;  b)+c)+d) Monge-Amp\`ere, transition and K\"ahler loss on validation data; e) $\sigma$-measure f) volume and g) $\mathcal{R}$-measure on test data; h) the linear relationship between improvement in $\sigma$-measure and $\mathcal{R}$-measure.\label{fig:fermat}}
	\end{figure}

	\paragraph{Neural Networks}
	
	The metric tensor $g$ is a $(3,3)$ hermitian complex tensor. The obvious ansatz is to let a neural network (NN) freely predict nine real numbers, $\gpred = g_{\text{NN}}$, such that we have a mapping $f: \mathbb{R}^{2 \sum n_i +1} \rightarrow \mathbb{R}^{9}$. As we will show in the next section, this is not the most efficient way to approach this problem, and one should modify the prediction of the NN with physical intuition. Equation~\eqref{eq:MA} shows that the unique Ricci-flat metric on $X$ is given by an exact correction to some reference K\"ahler metric $\gFSX$, which we may construct by pullback from the ambient space $\mathcal{A} =  \Pi_{i=1}^r \mathbb{P}^{n_i}$.\footnote{$\mathcal{A}$ admits a metric, $\gFS$
		, known as a Fubini-Study metric. The pullback $\gFSX = I^i_a (\gFS)_{i\bar{j}} I^{\bar{j}}_{\bar{b}}$ (where $I^i_a = \partial z_i/\partial x_a$, and $z_i$ and $x_a$ are coordinates on $\mathcal{A}$ and $X$) yields a globally well-defined K\"ahler metric  on $X$.} We then define NNs that learn corrections to $\gFSX$ such that the resulting metric $\gpred$ is Ricci-flat: addition $\gpred = \gFSX + g_{\text{NN}}$, element-wise multiplication $\gpred =\gFSX + \gFSX \odot g_{\text{NN}}$, matrix product $\gpred = \gFSX + \gFSX \cdot g_{\text{NN}}$, and Monge-Amp\`ere type $\gpred = \gFSX + \partial \bar{\partial} \phi$. 
	
	
	\paragraph{Loss contributions}
	
	We define five loss functions that ensure that the numerical metric satisfies the Monge-Amp\`ere equation~\eqref{eq:MA}, is K\"ahler, i.e. d$J = 0$, is well defined over patch transitions, has vanishing Ricci curvature~\eqref{eq:ricciscalar}, and results in the overall volume as the reference metric $\gFSX$ (this is a necessary condition to preserve the K\"ahler class). The total training loss is thus
	\begin{align}
		\label{eq:loss}
		\cL &= \alpha_1 \cL_{\text{MA}} + \alpha_2 \cL_{\text{dJ}} + \alpha_3 \cL_{\text{transition}} + \alpha_4 \cL_{\text{Ricci}} + \alpha_5 \cL_{\text{vol-K}}
	\end{align}
	where $\alpha_i$ are hyperparameters (by default set to $\alpha_i = 1.0$), and
	\begin{align}
		\label{eq:mariloss}
		&\cL_{\text{MA}} = \left|\left| 1 - \frac{1}{\kappa} \frac{\det \gpred}{\Omega \wedge \bar\Omega}\right|\right|_n \text{,} \qquad     \cL_{\text{Ricci}} = ||R||_n = \left|\left|\partial \bar\partial \ln{\det\gpred}\right|\right|_n \; ,
		\\
		\label{eq:kloss}
		& \cL_{\text{dJ}} = \sum_{ijk}  \left|\left|\Re{c_{ijk}}\right|\right|_n +  \left|\left|\Im{c_{ijk}}\right|\right|_n,
		\quad \text{with } c_{ijk} = g_{i\bar{j},k} - g_{k\bar{j},i} \quad \text{ and } g_{i\bar{j},k} = \partial_k g_{i\bar{j}} \; 
		\\
		\label{eq:tvolkloss}
		&\cL_{\text{transition}} = \frac{1}{d} \sum_{(s,t)} \left|\left|\gpred^{(t)} - T_{(s,t)} \cdot \gpred^{(s)} \cdot T^\dagger_{(s,t)}\right|\right|_n \; \text{ and } \;     \cL_{\text{vol-K}} = \left|\left| \int \det \gFS -  \int \det \gpred \right|\right|_n \; .
	\end{align}
	Here the subscript $n$ denote the $L_n$ norms (default is $n = 1$ for all but $\cL_{\text{dJ}}$ which has $n=2$), $T_{(s,t)} = \partial \Vec{x}^{(s)} / \partial \Vec{x}^{(t)} $ denote transition matrices between patches $s$ and $t$, $d$ is the number of patch transitions, and the integral is taken over mini-batches. By virtue of the Calabi-Yau theorem, $\cL_{\text{MA}}$ is a surrogate loss (which does not involve costly derivatives) for the Ricci-loss $\cL_{\text{Ricci}}$; hence the latter is disabled by default. Moreover, the $\phi$-model is by construction K\"ahler with fixed size moduli, and thus comes with a disabled $\cL_{\text{dJ}}$ and $\cL_{\text{vol-K}}$. It does requires two derivatives with respect to the input parameters instead, which makes it comparable in speed to the other NNs. The derivatives with respect to the input coordinates are computed with TensorFlow's automatic differentiation, which works reliable even when training the $\phi-$model against the Ricci-scalar with a total of five nested gradient tapes.
	
	\begin{figure}
		\begin{minipage}[c]{0.75\textwidth}
			\centering
			\includegraphics[width=1.\textwidth]{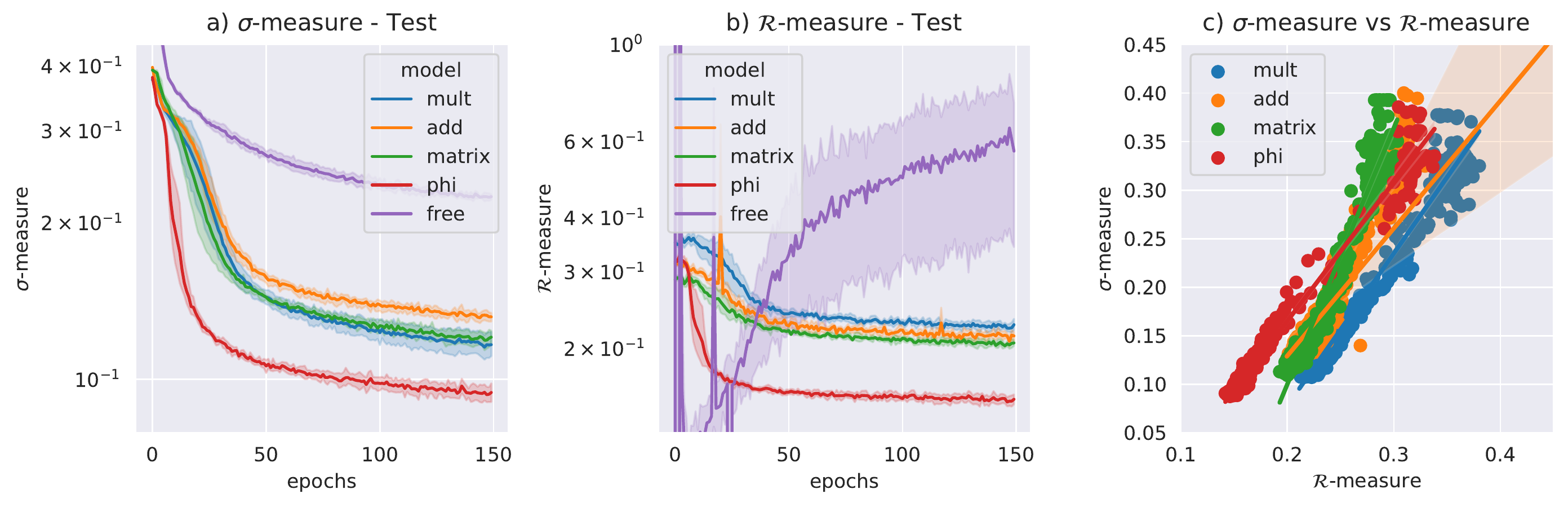}
		\end{minipage}\hfill
		\begin{minipage}[c]{0.25\textwidth}
			\tiny
			d) Slope computations.
			\label{tab:slope}
			\centering
			\begin{tabular}{@{}c|cc@{}}
				\toprule
				charges & $\phi$-model & true \\ \midrule
				(1,-1) & $0.06 \pm 0.05$  & 0 \\[4pt]
				(-1,1) & $-0.06 \pm 0.05$  & 0 \\[4pt]
				(1,1) & $17.96 \pm 0.06$ & 18 \\[4pt]
				(3,2) & $44.94 \pm 0.16$  & 45 \\[4pt]
				(3,-2) & $9.14 \pm 0.15$  & 9 \\[4pt]
				\bottomrule
			\end{tabular}
		\end{minipage}
		\caption{Bicubic experiments: a) $\sigma$-measure b) $\mathcal{R}$-measure and c) the linear relationship between improvement in $\sigma$-measure and $\mathcal{R}$-measure and d) mean slope (with one standard deviation as error) computations for five line bundles on test data.\label{fig:bicubic}}
	\end{figure}
	
	\section{Experiments}
	\label{sec:experiments}

	\paragraph{Fermat Quintic}
	
	To illustrate our package and benchmark it against previous results we first consider the Fermat Quintic, which is defined as the zero locus of $Q = \sum_{i=0}^4 z_i^5$ in $\mathcal{A} = \mathbb{P}^4$.
	Figure~\ref{fig:fermat} shows the results for each of the five model ans\"atze introduced in the previous section averaged over five runs. Training is done with an Adam optimizer on a fully connected feed-forward network with three hidden layers of 64 units each and GELU activation function for 100 epochs. Since we only want to learn a correction to the known Fubini-Study metric, we initialize the NN weights with small Gaussian variance $\mathcal{N}(0,0.01)$ ($\mathcal{N}(0,1)$ for the free network). The training set contains 198.000 points with a 0.1 validation split. We performed hyperparameter tuning via box searches for the number of hidden layers and activation function.
	
	As seen in Figure~\ref{fig:fermat}, the two multiplication networks and the $\phi$-model consistently arrive at the Ricci-flat metric.  We evaluate the  
	performance of the NN on the established benchmarks~\citep{Douglas:2006hz,Anderson:2010ke} of sigma and Ricci measures, $(\sigma,\mathcal{R})$,
	with an additional test set of 22.000 points in subplots e) and g). The $\phi$-models with lowest validation loss reach, within a few hours on a laptop, a mean accuracy of $\sigma = 0.0086$ and $\mathcal{R} = 0.076$. This is on par with $k=20$ in Donaldson algorithm~\citep{Anderson:2010ke} with a training time of 35 years on $4.6$E$8$ points~\citep{Ashmore:2019wzb}.  
	We are the first to systematically study $\mathcal{R}$, and demonstrate, in subplot h), a linear relation, $\sigma \approx 0.06 \mathcal{R}$, between optimization of the surrogate Monge-Amp\`ere equation~\eqref{eq:MA} and decrease in Ricci measure $\mathcal{R}$. Finally, subplot f) shows that the normalized volume of the manifold is constant, and hence that we do not change the K\"ahler class during training. 
	
	\paragraph{Bicubic}
	As a second example we consider the bicubic manifold $X$ given by a homogeneous degree (3,3) polynomial in $\mathcal{A} = \mathbb{P}^2 \times \mathbb{P}^2$ and size moduli, $t_1 = 1 = t_2$. For phenomenological reasons, and also to improve performance, we choose a member of this family which is invariant under a $\mathbb{Z}_3$ symmetry in its shape moduli.
	Figure~\ref{fig:bicubic} shows the averaged results of five experiments using the previously described experimental setup. As for the quintic, the three ``multiplicative'' NNs that are learning corrections relative to the value of the reference metric $\gFSX$ outperform the free and additive NNs, whose output is not scaled with the reference metric and varies over several orders of magnitude.  
	The $\phi$-model is particularly successful and the linear relation persists, $\sigma \approx 1.4 \mathcal{R} - 0.1$.
	
	For CYs with multiple size moduli $t_i$, an important consistency check of the $t_i$-dependence of the NN is given by the so-called slope, which is defined as $\mu =  \int \sqrt{g} g^{a \bar{b}} F_{a \bar{b}}$ for a vector bundle $V$ with  curvature $F$. Clearly, this slope depends not only on the CY volume, encoded in $\sqrt{g}$, but on the full metric tensor that is approximated by the NN. Crucially, $\mu$ may also be computed exactly using topological data of $(X,V)$, and can in fact be shown to vanish for stable, holomorphic $V$ \cite{donaldson,uhlenbeck-yau}.
	Subtable d) of Figure~\ref{fig:bicubic} shows that the slope computed using the $\phi$-model for various line bundles is in excellent  agreement with the exact, topological results, both for stable and non-stable bundles. 
	
	\section{Conclusion}
	\label{sec:conclusion}
	
	We have presented a computational library that learns Ricci-flat metrics for a wide range of CY manifolds.  The package comprises five conceptually different NN architectures, and we compare their efficacy in learning the metric. Two novel features compared to previous studies, are that the  numerical metrics can be computed for any value of the CY shape and size moduli, and the package allows for direct studies of the Ricci scalar.  Our package has a significantly lower entry barrier than previous methods. Moreover, using domain knowledge in point sampling and during training greatly improves the performance.  We emphasize that our results can be further improved by more careful hyperparameter optimization, more sample points, or simply longer training time. 
	
	\section*{Broader Impact}
	
	We have more efficient algorithms than before, which reduces the carbon footprint of numerical computations for phenomenological and mathematical studies of string theory. We democratize the learning of Calabi-Yau metrics by creating an open source package with low entry barrier, to the benefit of researchers that have smaller computational budgets. This may in part offset the positive environmental effect just mentioned, as energy consumption will grow with the number of users of the package. The package can be used to test swampland conjectures which are actively discussed in the string theory community. This might open up resources for other applications down the line.
	
	\begin{ack}
		RS and ML are funded in part by the Swedish Research Council (VR) under grant number 2020-03230. The work of FR is supported by startup funding from Northeastern University.
		Computations were in part enabled by resources provided by the Swedish National Infrastructure for Computing (SNIC) at the HPC cluster \emph{Tetralith}, partially funded by the Swedish Research Council through grant agreement no.\ 2018-05973, the hydra cluster of the Physics Department at the University of Oxford, and the computing cluster at CERN.
	\end{ack}
	
	\medskip
	
	\bibliographystyle{unsrturl}
	\bibliography{ref}
\end{document}